\documentclass[prl,aps,twocolumn,floats,nofootinbib,superscriptaddress]{revtex4-1}
\usepackage{braket}
\usepackage{appendix}

\usepackage{amsmath,amssymb,graphicx,psfrag}
\usepackage{amsmath} 
\usepackage[utf8]{inputenc}
\usepackage[colorlinks,linkcolor={blue},citecolor={blue},urlcolor={blue}]{hyperref}

\usepackage{amssymb}
\usepackage{algorithm}
\usepackage{babel}
\usepackage{algpseudocode}
\usepackage{bbm}
\usepackage[justification=justified]{caption}
\usepackage{caption}

\usepackage{graphicx}
\usepackage{subcaption}
\usepackage[font=small,labelfont=bf,
   justification=justified,
   format=plain]{caption} 
\usepackage{times}
\usepackage[margin=0.75in]{geometry}

\begin{document}

\preprint{APS/123-QI}

\title{Breaking Quantum Key Distributions under Quantum Switch-Based Attack }

\author{Sumit Nandi}
\email{sumit.enandi@gmail.com}
\affiliation{Purandarpur High School, Purandarpur, West Bengal 731129, India}


\author{Biswaranjan panda}
\email{biswaranjanpanda2002@gmail.com}
\affiliation{Centre for Quantum Engineering, Research and Education (CQuERE), TCG CREST, Kolkata, India }

\author{Pankaj Agrawal}
\email{pankaj.agrawal@tcgcrest.org}
\affiliation{Centre for Quantum Engineering, Research and Education (CQuERE), TCG CREST, Kolkata, India
}
\author{Arun Kumar Pati}
\email{patiqubit@gmail.com}
\affiliation{Synergy Quantum, Second Floor, Research and Innovation Park\\ 
Indian Institute of Technology Delhi, Hauz Khas, New Delhi, India}


\date{\today}

\begin{abstract}

Quantum key distribution (QKD) enables secure key sharing between distant parties, with several protocols proven resilient against conventional eavesdropping strategies. Here, we introduce a new attack scenario where an eavesdropper, Eve, exploits a quantum switch using the indefinite causal order to intercept and manipulate quantum communication channel. Using multiple metrics such as the information gain, mutual information, and Bell violation, we demonstrate that the presence of a quantum switch significantly compromises QKD security. Our results highlight a previously overlooked vulnerability, emphasizing the need for countermeasures against quantum-controlled adversarial strategies.

\end{abstract}

\maketitle

{\it Introduction}.--
Any secret communication strategies require the sharing of a key between communicating parties to encrypt and decrypt messages. Quantum key distribution (QKD) \cite{gisin_review} uses tenets of quantum mechanics for the generation of
a key. First, such a protocol was proposed by the pioneers Bennett and Brassard in $1984$ \cite{bb84} and is popularly known as BB84 protocol. Since then, many other QKD protocols for distributing secret keys over insecure quantum channel have been formulated \cite{ekert91,bbm92,vaidman95,acin07}. The striking difference between QKD and its classical counterpart lies in uncompromising security between the authenticated observers. Quantum mechanics underpins a protocol's security - any act of measurement by an eavesdropper disturbs a quantum system. Observation of detectable disturbances can lead to the detection of a secret
observer. In a realistic QKD scenario, a potential eavesdropper could follow any strategy allowed by the laws of physics to get knowledge of the secret keys. Some of the well-known QKD protocols have been shown to be secure under multiple attack scenarios \cite{gisin_review}.
 
In quantum key distribution protocol, two legitimate partners, traditionally known as Alice and Bob, are connected by a quantum channel and an authenticate classical channel. We also assume the presence of an unauthenticated person, Eve, whose main intention is to have optimal information about the secret message. Furthermore, Eve can have full control over the quantum channel between Alice and Bob. At the onset of the BB84 protocol the sender, Alice, prepares a qubit on a basis randomly chosen from a set of observables known to both Alice and Bob. Bob receives the qubit and measures on a basis randomly chosen from a set of observables. They publicly disclose encoding and decoding scheme via some classical channels. In the sifting procedure, they only retain those bit values for which their measurement basis are same.
Then comes the most crucial part of the protocol in which they publicly compare their bit values randomly chosen from the sifted keys, and estimate the fraction of keys for which their bit values are different known as quantum bit error rate (QBER), which bounds Eve’s information. If the error rate exceeds a critical threshold (approximately 15\%) \cite{gisin_review}, the protocol is aborted.  Since all quantum channels are inherently noisy, any practical QKD protocol must have non-zero QBER. In our present context, we assume that QBER of whatsoever amount is caused by Eve. Meanwhile, Eve with full supremacy over the quantum channel can eavesdrop on the protocol which is essentially a general kind of quantum operation on the qubits passing through the channel \cite{Huttner01121994,bruss98,Lutkehaus, gisin97}. Here, we focus on a specific eavesdropping strategy, called \emph{individual attack}, in which Eve captures the traversing qubits one after another and measures it. 
In the given paradigmatic situation, an optimal individual attack has been formulated in \cite{griffiths,scarani} to address the question like how much information Eve can have about Alice's key. The attack is optimal in the sense that for a given QBER eavesdropping strategy maximizes Eve's knowledge about Alice's key. In this letter, we revisit the attack strategy by letting Eve access to a more general kind of quantum operation assisted by a quantum switch.   

Quantum mechanics allows events to occur in no \emph{definite causal order}, and indefiniteness of ordering can be implemented by a quantum switch operation \cite{chiribella2013pra,brukner12}. In a quantum switch, a control-qubit controls the order of the operation of two channels, say $\mathcal{E}$ and $\mathcal{F}$. If the 
control qubit is in a superposed state, it leads to the superposition of the alternate order of the operation of these channels, in contrast to classical circuits where operations occur in a fixed causal order. This novel way of applying the operations holds promise for advancing quantum computing, communication, and metrology, by offering new ways to perform tasks that are impossible within the classical realm \cite{chiribella2009theoretical}. 
Quantum switch has demonstrated significant improvements in quantum communication protocols, particularly in cases with high noise levels \cite{ebler2018enhanced,
zhao2020quantum,akpati2020,grover_switch}. Experimental implementation of a quantum switch acting on arbitrary quantum channels has been realized recently, as discussed in \cite{rubino2017exp,caonatureexp,PhysRevLett.131.060803}.

In this letter, we introduce a novel framework of quantum switch enabled \emph{individual} eavesdropping strategy in a QKD protocol. In contrast to previously known attack strategies in which Eve was to intercept the incoming particles individually and perform a joint unitary operation with an ancillary qubit, our eavesdropping protocol relies on quantum switch operation on Eve's end. 
We show that the information gain achieved by Eve is significantly higher. We demonstrate that both the BB84 and E91 quantum key distribution (QKD) protocols are insecure under quantum switch-based attacks, thus, revealing a fundamental vulnerabilities in QKD. 
We also investigate the impact of the quantum switch application on mutual information and Bell inequality violation of the subsystems shared between Alice (Bob) and Eve. Finally, we look into the potential implication of our formalism by considering the symmetric individual attack which is a variant of an intercept-resend attack strategy.

{\it Quantum Switch enabled attack in QKD.}-- In a typical entanglement-based QKD protocol, a pair of entangled qubits in \emph{Bell state} is shared between Alice and Bob. Eve individually captures the qubits sent to Bob and uses an attack scheme which is basically a quantum operation on the joint system of the captured qubit and Eve's ancillary qubit. As the state space is two-dimensional, it suffices to assume that the ancillary system is also a two-dimensional system. Eve interacts with her measuring device separately on the captured particles before it reach Bob. Eve can use the same tactics of intercept-and-resend in the case of prepare-and-measure type of QKD protocols, like BB84. To gain maximal information of the captured qubit, Eve can implement the Scarani-Gisin joint unitary operation $U_{SG}$  \cite{scarani} on the ancillary and Bob's qubit. The operation of $U_{SG}$ on the joint product bases is described as 

\begin{align}\label{sunitary}
U_{SG}\ket{00} &= \ket{00}, \\
U_{SG}\ket{10} &= \cos\phi \ket{10} + \sin\phi \ket{01},
\end{align}
where $\phi$ $\in [0, \pi/2]$ is the strength of Eve's attack and it necessarily describes the QBER of the protocol. It is easy to check from the above equations that Eve's intervention introduces an error, so-called QBER, in the sifted key which is given by $\frac{\sin^2\phi}{2}$. Now, the final state of Alice, Bob, and Eve is given by
\begin{equation}\label{com_state}
\rho_{SG} ^{ABE}=I\otimes U_{SG}(|\tilde{\Psi^+}\rangle\langle\tilde{\Psi^+}|)I\otimes U^{\dagger}_{SG},
\end{equation}
where $|\tilde{\Psi^+}\rangle=\ket{\Phi^+}\otimes |0\rangle$ and $\ket{\Phi^+}=\frac{1}{\sqrt 2}(\ket{00}+\ket{11})$.

In quantum mechanics it is possible to have superposition of the ordering of events, known as the indefinite causal order, thereby, providing a more general framework of quantum operation. 
A quantum switch exploits superposition of casual ordering of quantum operations \cite{chiribella2013pra,brukner12}. It is composed of two quantum channels
$\{\mathcal{E},\mathcal{F}\}$, the order is controlled by an ancillary qubit. If the
state of the control qubit is $\ket{0}$, $\mathcal{F}$ is applied before $\mathcal{E}$, but if it is $\ket{1}$, then $\mathcal{E}$ is applied before $\mathcal{F}$. However, if the control qubit is in a superposition of $\ket{0}$ and $\ket{1}$, the quantum switch will create a superposition of the two alternative orders.
The action of the quantum operation $\mathcal{E}$, and $\mathcal{F}$ on a density operator $\rho$ can be expressed as
\begin{eqnarray}
\mathcal{E}(\rho)&=&\sum_i E_i\rho E_i^\dagger,\\
\mathcal{F}(\rho)&=&\sum_i F_i\rho F_i^\dagger.
\end{eqnarray} 
Here $\{E_i\}$, $\{F_j\}$ are operator elements of the corresponding quantum operations. Sequential operations as mentioned give rise to the following operations:
\begin{eqnarray}
\mathcal{E}\circ\mathcal{F}(\rho)&=&\sum_{i,j}E_jF_i\rho F_i^\dagger E_j^\dagger,\\
\mathcal{F}\circ\mathcal{E}(\rho)&=&\sum_{i,j}F_jE_i\rho E_i^\dagger F_j^\dagger.
\end{eqnarray} 
However, if choose the control qubit to be in a superposition $| \omega \rangle =\frac{1}{\sqrt2}(|0\rangle+|1\rangle)$, then the resultant quantum switch operation is expressed as

\begin{equation}\label{srho}
S(\rho \otimes \omega )=\sum_{i,j}M_{ij}(\rho \otimes \omega)M_{ij}^{\dagger},
\end{equation}
where $M_{ij}=E_iF_j\otimes|0\rangle\langle0|+F_jE_i\otimes|1\rangle\langle 1|$ and $\omega = | \omega \rangle \langle \omega |$.
It can also be verified that $M_{ij}$ satisfies $\sum_{i,j}M_{ij}M_{ij}^\dagger=I$.  

\begin{figure}[htbp]
    \centering    \includegraphics[width=0.9\linewidth]{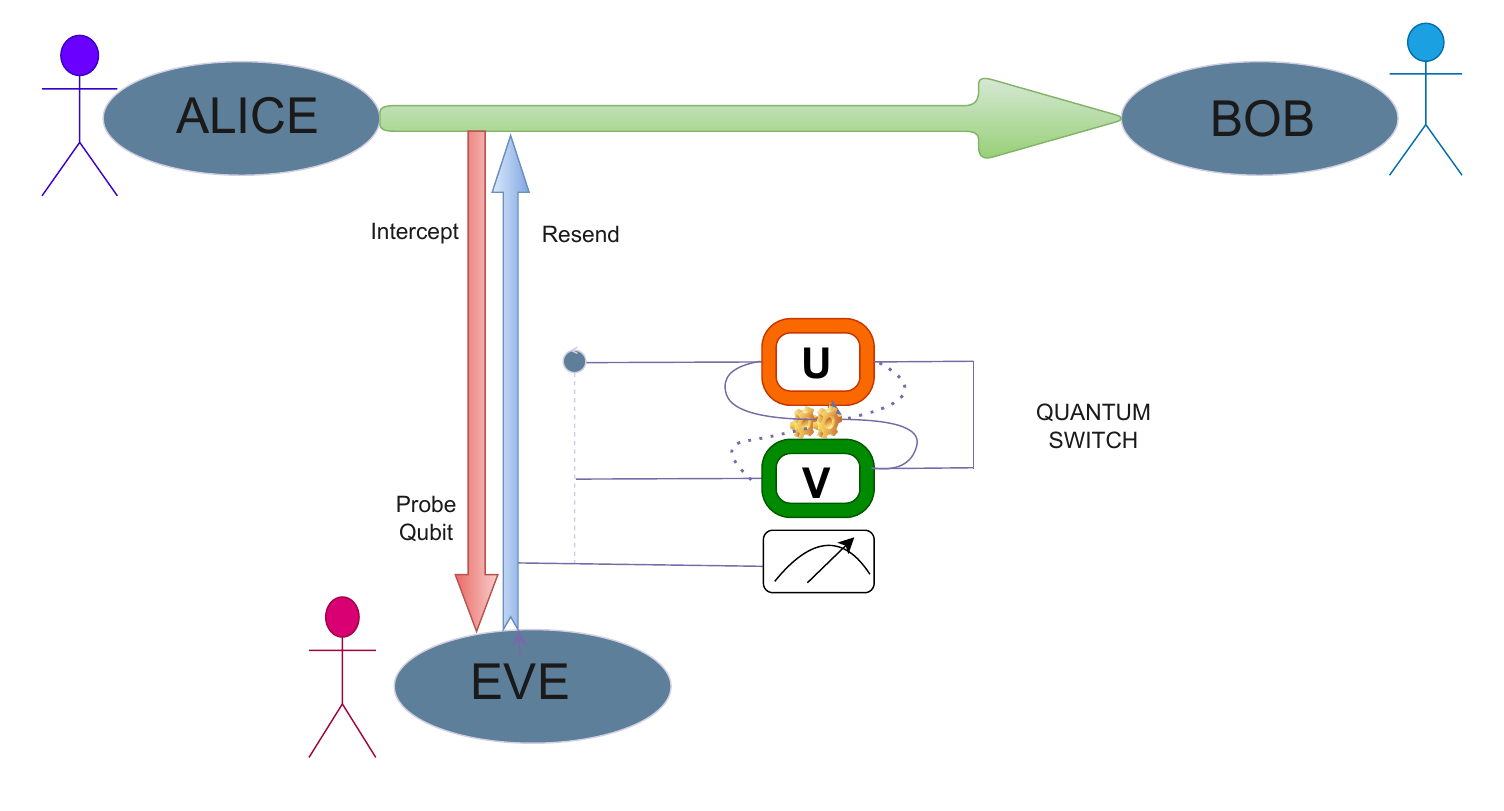}
    \caption{Schematic diagram of quantum switch enabled attack on a quantum communication channel.}
    \label{protocol}
\end{figure}

In the quantum switch-based eavesdropping strategy, Eve has access to another joint unitary, \emph{possibly} a number of such unitaries, and a quantum device that can perform quantum switch operation upon the incoming particles in a larger Hilbert space, illustrated in Fig. (\ref{protocol}). This approach can be viewed as an intercept-and-resend attack supplemented with quantum switch operation. To be specific, Eve applies quantum switch operation on Alice's qubits before resending them to Bob. Subsequently, Eve measures her probe qubit to extract information about the transmitted state in an optimized measurement setting. In what follows, quantum switch-enabled attack differs fundamentally from usual individual attacks. Before we present our main results, let us consider a quantum switch operation consisting of two unitaries $U$ and $V$ in indefinite causal order that act on a state $\rho$. 
The final state of the system of interest after the
measurement on the control qubit in the $\{ |+ \rangle \} $
basis can be expressed as \cite{indranilda}

\begin{equation}\label{Omega}
S(\rho)= \frac{\Lambda\rho\Lambda^\dagger }{ \mathrm{Tr} (\Lambda\rho\Lambda^\dagger)},
\end{equation}
where $\Lambda$ is given by 
$\Lambda=\frac{1}{2}(UV+VU)$.

It shows that the resultant switch operation does not follow the standard sequential type of casual sequence of operation. Subsequent switch operation of Eve
results in the following tripartite state shared between Alice, Bob and Eve

\begin{equation}\label{com_state_switch}
\tilde{\rho}_{SW}^{ABE}=\frac{I\otimes \Lambda(|\tilde{\Psi^+}\rangle\langle\tilde{\Psi^+}|)
I\otimes \Lambda^+}{\mathrm{Tr}\big(I\otimes \Lambda(|\tilde{\Psi^+}\rangle\langle\tilde{\Psi^+}|)
I\otimes \Lambda^+\big)}.
\end{equation}
In the sequel, we provide the security analysis of QKD with quantum switch-enabled attack. 
        
%

{\it Information Gain.}--The interaction between Eve's probe and the target qubit sent to Bob fully determines her accessible information. In order to extract that information, Eve measures her system with two different measurement settings $x_i$ $(i=1,2)$ parametrized by $\alpha_i$ 
\begin{eqnarray}\label{settings}
\ket{m_{x_i+}}=\cos\frac{\theta_i}{2}\ket{0}+\sin\frac{\theta_i}{2}\ket{1},\nonumber\\
\ket{m_{x_i-}}=\sin\frac{\theta_i}{2}\ket{0}-\cos\frac{\theta_i}{2}\ket{1}.
\end{eqnarray}
The probabilities of such general measurements on the joint state $\rho_{AE}$ can be obtained as
\begin{equation}
    P_{\lambda|x_i} = \mathrm{Tr}(\rho_{AE}.|\tilde{m}_{x_i+(-)}\rangle\langle \tilde{m}_{x_i+(-)}|),
\end{equation}
where $\lambda\in\{+,-\}$, and $|\tilde{m}_{x_i+(-)}\rangle\langle \tilde{m}_{x_i+(-)}|=I\otimes|m_{x_i+(-)}\rangle\langle m_{x_i+(-)}|$.  Probabilities $ P_{\lambda|x_i}$ encapsulates a great deal of information about the qubit sent to Bob. Let Alice send qubits completely randomly among two bases, then, the probability that Eve observes the outcome \(\lambda\) is given by 
\begin{equation}
q_\lambda = \frac{1}{2}\sum_{x_i} P_{\lambda|x_i}.
\end{equation}
%
Then the posterior probability (or likelihood) that Eve observes outcome \(\lambda\) 
 is given by Bayes’ theorem $Q_{i\lambda} = \frac{P_{\lambda|x_i} }{2q_\lambda}$. 
A convenient measure of Eve’s information gain is given by~\cite{ref13}
\begin{equation}
G_\lambda = \left| Q_{x\lambda} - Q_{y\lambda} \right|.
\end{equation}
The amount of information that Eve can gather on the average is given by   
\begin{equation}
\sum_\lambda q_\lambda G_\lambda = \sum_\lambda \left| P_{\lambda|x} p_x - P_{\lambda|y} p_y \right|.
\end{equation}
For completely random measurement settings, we obtain
%

\begin{equation}\label{info_gain}
G = \frac{1}{2} \sum_\lambda \left| P_{\lambda|x} - P_{\lambda|y} \right|.  \end{equation}
In the light of the above discussion, we present one of the main results by showing that the quantum switch-enabled attack scheme provides a larger value of $G$.
By using Eq. (\ref{info_gain}), we calculate information gain by Eve in two different cases as follows. Firstly, we consider the joint tripartite state given by Eq.~(\ref{com_state}) and obtain $\rho_{AE}$ by tracing over the subsystem $B$. 
We use Eq.~(\ref{info_gain}) to calculate information gain denoted as $G_{SG}$ 
\begin{equation}\label{gws}
    G_{SG} =  0.25\cos^2\phi.
\end{equation}
To compute the quantities $P_{\lambda|\ell}$ ($\ell=x,y$), we use $\theta_1 = 0$, $\theta_2 = \pi/2$, respectively, in measurement settings given by  Eq. (\ref{settings}).
In order to obtain information gain in the later case, i.e., invoking quantum switch operation, one needs another joint unitary operation. We take $U_{SG}$ and
Pauli XZ-gate \cite{nielsen} to construct quantum switch operation and obtain

\begin{equation}
\Lambda(\phi)=\frac{1}{2}\big(U_{SG}.XZ+XZ.U_{SG}\big),
\end{equation}
where $\phi$ characterizes the strength of the switch operation. We obtain $\tilde{\rho}_{AE}$ by taking partial over subsystem $B$ in Eq.(\ref{com_state_switch}), and finally using  Eq.~(\ref{info_gain}) corresponding expression of information gain denoted as $G_{SW}$ turns out to be $G_{SW}=0.25\cos\phi $.
The ratio of information gain with and without the quantum switch operation, thus, obtained as 
\begin{equation}\label{xzgate}
 \frac{G_{SW}}{G_{SG}}=\sec\phi>1,
\end{equation}
where $\phi$ $\in$ $(0, \pi/2)$. It shows that quantum switch-enabled attack scheme can be more advantageous to Eve toward her ultimate intention of acquiring more information. 
We note that the choice of unitary operations required to construct quantum switch plays very important role in eavesdropping. For illustration, we consider SWAP gate \cite{nielsen} and $U_{SG}$ to devise switch operation and find the following expression for information gain $G^\prime_{WS}$

\begin{equation}\label{gsw}
G^\prime_{SW}=\left| \frac{1}{\cos (2 \phi )+3}-\frac{1}{4}\right|
\end{equation}  
Evidently, the ratio $\frac{G_{SW}}{G_{SG}}>1$ \emph{iff} $\phi>\frac{\pi}{4}$.

\vspace{0.1cm}

\begin{figure}[ht!]
    \centering
    
    \begin{subfigure}[b]{0.49\linewidth}
        \centering
        \includegraphics[width=\linewidth]{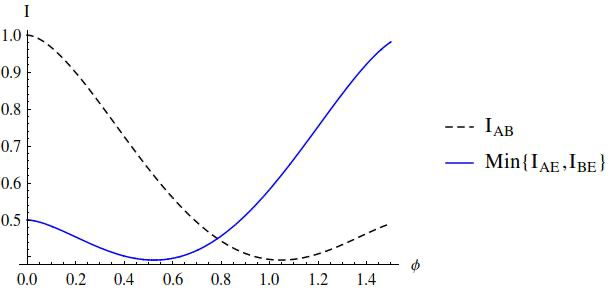}
        \caption{}
        \label{fig:subplot1}
    \end{subfigure}
    \hfill
    \begin{subfigure}[b]{0.49\linewidth}
        \centering
        \includegraphics[width=\linewidth]{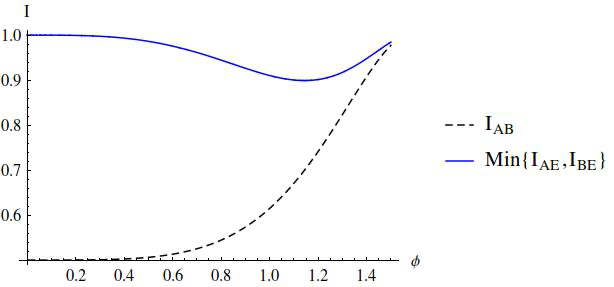}
        \caption{}
        \label{fig:subplot2}
    \end{subfigure}
    
    \caption{Fig. 2(a) illustrates the plot between \( \phi \) and mutual information ($I$) of different subsystems of $\rho_{SG} ^{ABE}$. The protocol remains secure when \( \phi \in [0, \pi/4] \), as it satisfies the condition given by Eq. (\ref{bound}). Fig. 2(b) compares \( I_{AB}= I(A:B) \) with \( \min \{ I_{AE}= I(A:E), I_{BE}= I(B:E) \} \) and it shows that \( \min \{ I(A:E), I(B:E) \} \) violates the security condition for the entire range of the parameter $\phi$}
    
    \label{fig:compare}
\end{figure}

{\it Mutual Information.}-- Another metric that we can use to assess the severity of the attack is to compute mutual information for Alice vs Bob $I_{AB} = I(A:B)$ and Alice-Bob vs Eve $I(AB:E)$. If $I(AB:E)$ exceeds $I(A:B)$, the security of the protocol has been seriously compromised.
 Mutual information is defined as
\begin{equation}\label{mutual_inf}
I(X:Y)= H(X) - H(X|Y),
\end{equation}
where $H(\{p_i\})$ is the Shannon entropy of the distribution $\{p_i\}$. We obtain density matrices \(\rho_{AB}\), \(\rho_{BE}\), and \(\rho_{AE}\) by tracing over appropriate subsystem of the joint tripartite state $\rho_{ABE}$ and $\tilde{\rho}_{ABE}$ given by Eq. (\ref{com_state}) and Eq. (\ref{com_state_switch}), respectively. In this case, we use $U_{SG}$ and $SWAP$-gate to construct the quantum switch operation. We explicitly provide $\tilde{\rho}_{ABE}$ in Appendix-1. We then evaluate mutual information between $A$, $B$, and $A$($B$), $E$ using the measurement settings given in Eq. (\ref{settings}), and plot the corresponding quantities in Fig. ({\ref{fig:compare}}) with respect to the parameter $\phi$. 
 
The plot in Fig. ({\ref{fig:subplot1}}) is obtained using the attack scenario as prescribed in \cite{scarani}. It suggests $\text{Min}\{I(A:E),(I(B:E)\}$ and $I(A:B)$ intersects at $\phi=\frac{\pi}{4}$. Attack as such had been shown to be very important in the context of the security of the QKD protocol. It provides Eve with optimal knowledge about the test qubit for a given QBER. However, it was shown in \cite{Bennett1992,Privecy_amp} that secret key extraction is indeed possible whenever Bob has more information on Alice's system than Eve, i.e., 
if the following condition holds 

\begin{equation}\label{bound}
I(A:B)> \mathrm{Min} \{I(A:E),(B:E)\}.
\textit{}\end{equation} 
Alice and Bob can continue the protocol by error correction and privacy amplification method. The inequality Eq. (\ref{bound}) was also shown to be necessary for one-way communication.
The plot in the right panel of Fig. (\ref{fig:subplot2}) is for Eve's attack with
the quantum switch using $U_{SG}$ and $SWAP$.
This plot has an interesting feature that is consequential and potentially far-reaching. The plot shows that $\textrm{Min}\{I(A:E),I(B:E)\}$ surpasses $I(A:B)$ for the entire range of the parameter $\phi$. Evidently, $\textrm{Min}\{I(A:E),(I(B:E)\}$ is also larger for \emph{very small} value of the QBER. In a QKD protocol, Alice and Bob generally abort the protocol if they obtain large QBER. Our eavesdropping strategy induces less noise to gain more information. This feature is very distinctive as it enables Eve to gain more knowledge of the secret key.

It should be noted that the severity of the attack depends on the choice of unitaries used to devise a quantum switch operation. We have also considered a switch operation by combining $U_{SG}$ and $CNOT$, $U_{SG}$ and $XZ$-gate. In both cases, the bound given by Eq.~(\ref{bound}) is violated for certain values of the parameter $\phi$. This bound has a significant physical interpretation: if \(I(A:B)\) dominates over \(I(A:E)\) and \(I(B:E)\) within a specific range of \(\phi\) values, the quantum protocol is considered secure. Otherwise, a security alert is warranted. 
In the case of the quantum switch attack, this bound is violated for all \(\phi \in [0, \pi/2)\), indicating that the security of the protocol is compromised.
In the next, we discuss the impact of the quantum switch-enabled attack on the optimal values of Bell inequality violation of different subsystems and its further implications.

\begin{figure}
\includegraphics[width=1.0\linewidth]
{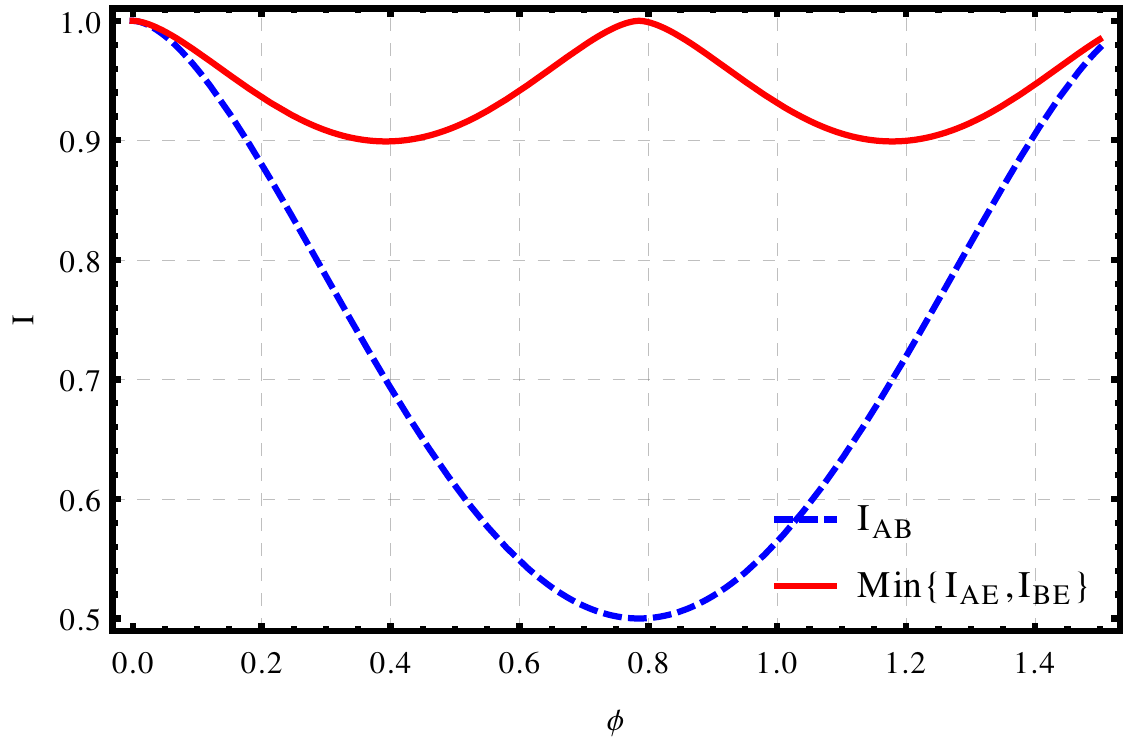}
\caption{The plot illustrates that $\min \ I(A:E), I(B:E) \}$ always violets the security condition as given by Eq. (\ref{bound}).
}
\label{scarani_bb84}
\end{figure}

\vspace{0.2cm}
{\it Violation of Bell inequalities}.--
Let us now investigate the situation by evaluating  Bell inequalities for $\rho_{AB}$, $\rho_{AE}$, and $\rho_{BE}$. The Bell inequality is given in terms of
\begin{equation}
\mathcal{B} = A_1 \otimes (B_1 + B_2) +  A_2 \otimes (B_1 - B_2),   
\end{equation}
where $A_1, A_2$ and $B_1, B_2$ are dichotomic
observables that can take the values $\{-1,1\}$.
 A state violates local realism, if $|\langle\mathcal{B}\rangle| > 2$, where $\langle.\rangle$ denotes the average taken over multiple rounds of measurements. Such states have quantum entanglement.
In order to investigate in detail, we obtain the maximum value of the Bell observable $\mathcal{B}$ for the density matrices $\rho_{AB}$, $\rho_{AE}$, and $\rho_{BE}$ obtained by tracing over appropriate subsystem from the joint state given in Eq.(\ref{joint_state}). Using the results in Horodecki \textit{et al.}~\cite{horodecki1995}, corresponding values of $\langle \mathcal{B} \rangle$ for the density matrices $\rho_{AB}$ and $\rho_{BE}$ vanish, and for $\rho_{AE}$ it turns out to be
 
\begin{eqnarray}
\langle\mathcal{B}_{AE} \rangle =2\sqrt{
 1 +\frac{16 \cos^4\phi}{(1 + \cos^2\phi)^4}}.
 \label{bellbound}
\end{eqnarray}
Clearly $\langle\mathcal{B}_{AE} \rangle > 2$ except for $\phi = {\pi \over 2}$. It has a deep consequence in the context of the security of the shared key between Alice and Bob. The violation of the inequality as in Eq.(\ref{bellbound}) does not create a problem as long as certain other conditions are satisfied. For example, if $\rho_{AB}$ violates Bell inequality then it would still be possible to establish secret key with larger numbers of shared states and following the privacy amplification procedure. But, here we encounter a more vulnerable situation, since, $\rho_{AB}$ does not violate Bell inequality at all. It completely rules out the possibility of carrying out the protocol with privacy amplification \emph{only}. Thus, the imposition of superposition in the ordering of Eve's interaction with the target qubit seems to be more advantageous for her. Thus, a quantum switch-based attack can make the entanglement based QKD protocols insecure.

Using mutual information metric, we can also see serious security risk to Prepare-and-measure protocols like BB84.
In this case, we consider two different attack strategies. In each case we consider two scenarios, one with 
and another without a quantum switch. In both cases, we find that a quantum switch based attack seriously 
compromises the protocol security.

{\it Intercept-and-resend attack}.-- We compare two cases: (i) when Eve uses only
$U_{SG}$ or (ii) a quantum switch using $U_{SG}$ and $SWAP$ during the intercept-and-resend attack.
It turns out that the BB84 protocol is highly vulnerable when Alice prepares the state and transmits it to Bob, and Eve uses a quantum switch to gain the information. In this case, the joint state
$\ket{\tilde{\Psi}}_{ABE}$ is given by  

\begin{equation}\label{joint_state}
\ket{\tilde{\Psi}}_{ABE}=\frac{1}{\sqrt{1+\cos^2\phi}}\big(\ket{000}+\cos
\phi\ket{101}).
\end{equation}
As we see that this state is biseparable.
 Alice's qubit is entangled with Eve's qubit,
 but not with that of Bob's. So there will be a Bell violation by $\rho_{AE}$, but not $\rho_{AB}$.
 As before, we obtain $\rho_{XY}$ from $\ket{\tilde{\Psi}}_{ABE}$ to evaluate mutual information $I(A:B)$, and $I(A(B):E)$. 
 In Fig.(\ref{scarani_bb84}), we plot mutual information of the subsystems with respect to the parameter $\phi$. This shows a similar trend as we obtained earlier, i.e., Eve is able to gain significant information about the secret key for very low QBER. Nevertheless, Alice and Bob can extract a secret key by using the purification
of entanglement. They may still be able to extract a secret key by a procedure, known as quantum privacy amplification in which the optimal limit of QBER for BB84 protocol is allowed up to $25\%$. However, our quantum switch-based attack strategy is more powerful, since, it rules out the possibility of privacy amplification of whatsoever kind. 
\\
Now, it would be very natural to ask if Eve gathers more information on Alice than Bob for the entire parameter range specified by $\phi$ which eventually introduces QBER, then what can be concluded about the authenticity of the protocol. It was shown that the protocol can still be considered by quantum privacy amplification \cite{qpa_des} if the quantum bit error rate is not high and $\rho_{AB}$ violates Bell inequality. We have already discussed that attack strategy based on quantum switch enables the eavesdropper to gain more mutual information \emph{even} at very low QBER.

{\em Symmetric individual attack}:
In this section, we discuss a particular variant of intercept and resend attack strategy to eavesdrop a typical BB84 protocol.
Here we deal with a specific kind of attack called symmetric attack in which Bob’s state is related to Alice’s state, \( \rho_A \), by a simple \emph{shrinking factor}, denoted as \( \alpha \). This factor captures the reduction in the Bloch vector introduced by the eavesdropping process. We express it as:  
\begin{eqnarray}
\rho_A(\vec{r}) = \frac{1}{2} \left( I + \vec{r} \cdot \vec{\sigma} \right) \;\; \text{and},\\
\rho_B(\vec{r}) = \frac{1}{2} \left( I + \alpha \, \vec{r} \cdot \vec{\sigma} \right),
\end{eqnarray}
where   \( \alpha \in [0, 1] \) is the shrinking factor that quantifies the level of disturbance.
From the above equation, it is well understood that such an attack is basis-independent and symmetric. 


%
To construct such a framework of symmetric attack, one needs unitary operation that realizes the following transformation on the joint bases state of Bob and Eve, respectively \cite{gisin_review}: 
\begin{equation}\label{gisin1}
    \mathcal{U}|\uparrow, 0\rangle = |\uparrow\rangle\otimes|\phi_{+}\rangle + |\downarrow\rangle\otimes|\theta_{+}\rangle,
\end{equation}
\begin{equation}\label{gisin2}
    \mathcal{U}|\downarrow, 0\rangle = |\downarrow\rangle\otimes|\phi_{-}\rangle + |\uparrow\rangle\otimes|\theta_{-}\rangle.
\end{equation}
Here, Bob and Eve's local spaces are spanned by the basis vectors $\{\ket{\uparrow},\ket{\downarrow}\}$ and $\{\ket{\phi_\pm},\ket{\theta_\pm}\}$ respectively.
%
%
$\{\ket{\phi_\pm},\ket{\theta_\pm}\}$ are not necessarily normalized.
Symmetry implies that \( |\langle\phi_+|\phi_{+}\rangle|^2 =|\langle\phi_-|\phi_{-}\rangle|^2 = F \) and \( |\langle\theta_+|\theta_{+}\rangle|^2 =|\langle\theta_-|\theta_{-}\rangle|^2 = D \). It can be checked that $F + D = 1$ and $\langle\phi_{+}|\theta_{-}\rangle + \langle\phi_{-}|\theta_{+}\rangle = 0$. Here $F$ and $D$ are very useful quantities and play a significant role in determining parity and disparity, respectively, of Bob's system with that of Alice's, and are called fidelity and QBER respectively. Now, we consider the mentioned attack strategy by letting Eve access an ancillary system and a suitable joint unitary operation to perform quantum switch-enabled eavesdropping. The security analysis of symmetric attack follows the same line as discussed in the preceding section. 
  Let \( P, Q \in \{A, B, E\} \), with \( P \neq Q \), represent two of the three parties involved. We evaluate \( I(P : Q) \) using Eq.(\ref{mutual_inf}) by considering two distinct probability statistics using the measurement settings given by Eq.(\ref{settings}). In order to obtain optimal attack by invoking quantum switch operation, we consider CNOT operation in this case. The joint normalized tripartite state $\ket{\chi^{SW}}_{ABE}$ is given by
\begin{equation}
\ket{\chi ^{SW}}_{ABE}=\cos\phi\ket{000}+
\frac{\sin\phi}{2}(\ket{101}+\ket{011}+\ket{100}+\ket{010}),
\end{equation}  
where $\phi\in[0,\frac{\pi}{2}]$ represents the strength of interaction $\mathcal{U}$ given by Eq.(\ref{gisin1}-\ref{gisin2}) of Eve's probe with the incoming systems.

It is noted that $I_{BE}$ dominates $I_{AB}$ for the entire range of QBER. Thus, the quantum switch makes symmetric individual attack more vulnerable by enabling Eve to gain optimal information of Alice's system \emph{even} within the safe periphery of $15\%$ QBER. 

\begin{figure}[ht!]
\begin{center}
\includegraphics[width=3.2in]{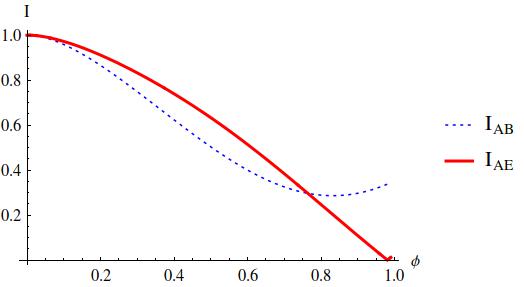}
\caption{The plot illustrates comparative behavior of the mutual information of different sub-systems of $|\chi^{SW}\rangle_{ABE} $ as a function of $\phi$. It indicates quantum switch-enabled symmetric individual attack is more vulnerable than the usual attack scenario.} \label{gisin_cnot}

\end{center}
\end{figure}

%

{\it Conclusions.}-- In this letter, we have presented a novel eavesdropping protocol in a typical QKD network. Our attack strategy successfully breaks the uncompromising security of the QKD protocols such as BB84, E91 and BBM92. In our attack scheme, Eve is allowed to make use of a quantum switch operation arising from indefinite causal order upon the individual state transmitted to Bob over a quantum channel. This kind of quantum switch-enabled attack outperforms other schemes based on the intercept and resend strategy. We have analytically shown that the information gain by Eve of the quantum switch-enabled attack is overwhelming. We have then illustrated distinctive features of our framework by obtaining mutual information about Eve with the authentic persons and presenting a detailed numerical analysis. It is shown that Eve's mutual information with Alice for a given QBER is significantly larger than the existing optimal attack strategies. 

Another noteworthy aspect of our strategy is explained by comparing the optimal Bell violation of Eve with Alice and Bob. Interestingly, it is found that the quantum switch induces maximal non-local correlation between Eve and Alice's subsystem while diminishing the correlation of Bob with Alice's subsystem. The novelty of our attack strategy lies in the fact that it completely discards the idea of establishing a secret key with one-way privacy amplification only. Lastly, the robustness of our quantum switch-enabled attack is presented by considering individual symmetric attack, a particular variant of intercept-and-resend attack strategy. It is shown that the quantum switch makes this particular attack more vulnerable by letting Eve gain more information about the secret key \emph{even} at very low QBER. 
Notably, there exist various other strategies of attacks in QKD protocol. Of particular interest is the class of attack where Eve makes operations on several qubits coherently, the so-called coherent attacks. It is in contrast to our strategy where it was assumed that Eve interacts with only one qubit at a given instant of time. In future, it will be worthwhile to investigate the impact of Eve's quantum switch-based attack in a coherent attack and other scenarios.   

Breaking quantum key distribution (QKD) would have profound implications for cybersecurity, as QKD is designed to provide unbreakable encryption based on the principles of quantum mechanics. Since quantum switch-based attack can severely compromise QKD protocols, it could render secure communication channels vulnerable, exposing sensitive data. Moreover, such a breach would challenge fundamental assumptions about quantum security, necessitating a reevaluation of cryptographic frameworks. This could lead to a race to develop new quantum-resistant security measures and redefine our approach to secure quantum communication.

\nocite{*}
\bibliography{sorsamp}

\begin{thebibliography}{32}%
\makeatletter
\providecommand \@ifxundefined [1]{%
 \@ifx{#1\undefined}
}%
\providecommand \@ifnum [1]{%
 \ifnum #1\expandafter \@firstoftwo
 \else \expandafter \@secondoftwo
 \fi
}%
\providecommand \@ifx [1]{%
 \ifx #1\expandafter \@firstoftwo
 \else \expandafter \@secondoftwo
 \fi
}%
\providecommand \natexlab [1]{#1}%
\providecommand \enquote  [1]{``#1''}%
\providecommand \bibnamefont  [1]{#1}%
\providecommand \bibfnamefont [1]{#1}%
\providecommand \citenamefont [1]{#1}%
\providecommand \href@noop [0]{\@secondoftwo}%
\providecommand \href [0]{\begingroup \@sanitize@url \@href}%
\providecommand \@href[1]{\@@startlink{#1}\@@href}%
\providecommand \@@href[1]{\endgroup#1\@@endlink}%
\providecommand \@sanitize@url [0]{\catcode `\\12\catcode `\$12\catcode `\&12\catcode `\#12\catcode `\^12\catcode `\_12\catcode `\%12\relax}%
\providecommand \@@startlink[1]{}%
\providecommand \@@endlink[0]{}%
\providecommand \url  [0]{\begingroup\@sanitize@url \@url }%
\providecommand \@url [1]{\endgroup\@href {#1}{\urlprefix }}%
\providecommand \urlprefix  [0]{URL }%
\providecommand \Eprint [0]{\href }%
\providecommand \doibase [0]{http://dx.doi.org/}%
\providecommand \selectlanguage [0]{\@gobble}%
\providecommand \bibinfo  [0]{\@secondoftwo}%
\providecommand \bibfield  [0]{\@secondoftwo}%
\providecommand \translation [1]{[#1]}%
\providecommand \BibitemOpen [0]{}%
\providecommand \bibitemStop [0]{}%
\providecommand \bibitemNoStop [0]{.\EOS\space}%
\providecommand \EOS [0]{\spacefactor3000\relax}%
\providecommand \BibitemShut  [1]{\csname bibitem#1\endcsname}%
\let\auto@bib@innerbib\@empty
\bibitem [{\citenamefont {Gisin}\ \emph {et~al.}(2002)\citenamefont {Gisin}, \citenamefont {Ribordy}, \citenamefont {Tittel},\ and\ \citenamefont {Zbinden}}]{gisin_review}%
  \BibitemOpen
  \bibfield  {author} {\bibinfo {author} {\bibfnamefont {N.}~\bibnamefont {Gisin}}, \bibinfo {author} {\bibfnamefont {G.}~\bibnamefont {Ribordy}}, \bibinfo {author} {\bibfnamefont {W.}~\bibnamefont {Tittel}}, \ and\ \bibinfo {author} {\bibfnamefont {H.}~\bibnamefont {Zbinden}},\ }\href {\doibase 10.1103/RevModPhys.74.145} {\bibfield  {journal} {\bibinfo  {journal} {Rev. Mod. Phys.}\ }\textbf {\bibinfo {volume} {74}},\ \bibinfo {pages} {145} (\bibinfo {year} {2002})}\BibitemShut {NoStop}%
\bibitem [{\citenamefont {Bennett}\ and\ \citenamefont {Brassard}(1984)}]{bb84}%
  \BibitemOpen
  \bibfield  {author} {\bibinfo {author} {\bibfnamefont {C.~H.}\ \bibnamefont {Bennett}}\ and\ \bibinfo {author} {\bibfnamefont {G.}~\bibnamefont {Brassard}},\ }in\ \href@noop {} {\emph {\bibinfo {booktitle} {Proceedings of IEEE International Conference on Computers, Systems, and Signal Processing}}}\ (\bibinfo  {publisher} {IEEE},\ \bibinfo {address} {Bangalore, India},\ \bibinfo {year} {1984})\ pp.\ \bibinfo {pages} {175--179}\BibitemShut {NoStop}%
\bibitem [{\citenamefont {Ekert}(1991)}]{ekert91}%
  \BibitemOpen
  \bibfield  {author} {\bibinfo {author} {\bibfnamefont {A.~K.}\ \bibnamefont {Ekert}},\ }\href@noop {} {\bibfield  {journal} {\bibinfo  {journal} {Physical Review Letters}\ }\textbf {\bibinfo {volume} {67}},\ \bibinfo {pages} {661} (\bibinfo {year} {1991})}\BibitemShut {NoStop}%
\bibitem [{\citenamefont {Bennett Charles~H.}(1992)}]{bbm92}%
  \BibitemOpen
  \bibfield  {author} {\bibinfo {author} {\bibfnamefont {M.~N.~D.}\ \bibnamefont {Bennett Charles~H.}, \bibfnamefont {Brassard~Gilles}},\ }\href {\doibase https://doi.org/10.1103/PhysRevLett.68.557} {\bibfield  {journal} {\bibinfo  {journal} {Physical Review Letters}\ }\textbf {\bibinfo {volume} {68}},\ \bibinfo {pages} {557} (\bibinfo {year} {1992})}\BibitemShut {NoStop}%
\bibitem [{\citenamefont {Goldenberg}\ and\ \citenamefont {Vaidman}(1995)}]{vaidman95}%
  \BibitemOpen
  \bibfield  {author} {\bibinfo {author} {\bibfnamefont {L.}~\bibnamefont {Goldenberg}}\ and\ \bibinfo {author} {\bibfnamefont {L.}~\bibnamefont {Vaidman}},\ }\href {\doibase https://doi.org/10.1103/PhysRevLett.75.1239} {\bibfield  {journal} {\bibinfo  {journal} {Phys. Rev. Lett.}\ }\textbf {\bibinfo {volume} {75}},\ \bibinfo {pages} {1239} (\bibinfo {year} {1995})}\BibitemShut {NoStop}%
\bibitem [{\citenamefont {Acín}\ \emph {et~al.}(2007)\citenamefont {Acín}, \citenamefont {Brunner}, \citenamefont {Gisin}, \citenamefont {Massar}, \citenamefont {Pironio},\ and\ \citenamefont {Scarani}}]{acin07}%
  \BibitemOpen
  \bibfield  {author} {\bibinfo {author} {\bibfnamefont {A.}~\bibnamefont {Acín}}, \bibinfo {author} {\bibfnamefont {N.}~\bibnamefont {Brunner}}, \bibinfo {author} {\bibfnamefont {N.}~\bibnamefont {Gisin}}, \bibinfo {author} {\bibfnamefont {S.}~\bibnamefont {Massar}}, \bibinfo {author} {\bibfnamefont {S.}~\bibnamefont {Pironio}}, \ and\ \bibinfo {author} {\bibfnamefont {V.}~\bibnamefont {Scarani}},\ }\href {\doibase https://doi.org/10.1103/PhysRevLett.98.230501} {\bibfield  {journal} {\bibinfo  {journal} {Phys. Rev. Lett.}\ }\textbf {\bibinfo {volume} {98}},\ \bibinfo {pages} {230501} (\bibinfo {year} {2007})}\BibitemShut {NoStop}%
\bibitem [{\citenamefont {Huttner}\ and\ \citenamefont {Ekert}(1994)}]{Huttner01121994}%
  \BibitemOpen
  \bibfield  {author} {\bibinfo {author} {\bibfnamefont {B.}~\bibnamefont {Huttner}}\ and\ \bibinfo {author} {\bibfnamefont {A.~K.}\ \bibnamefont {Ekert}},\ }\href {\doibase 10.1080/09500349414552301} {\bibfield  {journal} {\bibinfo  {journal} {Journal of Modern Optics}\ }\textbf {\bibinfo {volume} {41}},\ \bibinfo {pages} {2455} (\bibinfo {year} {1994})}\BibitemShut {NoStop}%
\bibitem [{\citenamefont {Brus}(1998)}]{bruss98}%
  \BibitemOpen
  \bibfield  {author} {\bibinfo {author} {\bibfnamefont {D.}~\bibnamefont {Brus}},\ }\href {\doibase https://doi.org/10.1103/PhysRevLett.81.3018} {\bibfield  {journal} {\bibinfo  {journal} {Phys. Rev. Lett.}\ }\textbf {\bibinfo {volume} {81}},\ \bibinfo {pages} {3018} (\bibinfo {year} {1998})}\BibitemShut {NoStop}%
\bibitem [{\citenamefont {L\"{u}tkenhaus}(1996)}]{Lutkehaus}%
  \BibitemOpen
  \bibfield  {author} {\bibinfo {author} {\bibfnamefont {N.}~\bibnamefont {L\"{u}tkenhaus}},\ }\href {\doibase https://doi.org/10.1103/PhysRevA.59.3301} {\bibfield  {journal} {\bibinfo  {journal} {Phys. Rev. A}\ }\textbf {\bibinfo {volume} {54}},\ \bibinfo {pages} {3301} (\bibinfo {year} {1996})}\BibitemShut {NoStop}%
\bibitem [{\citenamefont {Gisin~N.}(1997)}]{gisin97}%
  \BibitemOpen
  \bibfield  {author} {\bibinfo {author} {\bibfnamefont {H.~B.}\ \bibnamefont {Gisin~N.}},\ }\href {\doibase 10.1016/S0375-9601(97)00460-X} {\bibfield  {journal} {\bibinfo  {journal} {Phys. Lett. A}\ }\textbf {\bibinfo {volume} {228}},\ \bibinfo {pages} {3301} (\bibinfo {year} {1997})}\BibitemShut {NoStop}%
\bibitem [{\citenamefont {Fuchs}\ \emph {et~al.}(1997)\citenamefont {Fuchs}, \citenamefont {Gisin}, \citenamefont {Griffiths}, \citenamefont {Niu},\ and\ \citenamefont {Peres}}]{griffiths}%
  \BibitemOpen
  \bibfield  {author} {\bibinfo {author} {\bibfnamefont {C.~A.}\ \bibnamefont {Fuchs}}, \bibinfo {author} {\bibfnamefont {N.}~\bibnamefont {Gisin}}, \bibinfo {author} {\bibfnamefont {R.~B.}\ \bibnamefont {Griffiths}}, \bibinfo {author} {\bibfnamefont {C.-S.}\ \bibnamefont {Niu}}, \ and\ \bibinfo {author} {\bibfnamefont {A.}~\bibnamefont {Peres}},\ }\href {\doibase 10.1103/PhysRevA.56.1163} {\bibfield  {journal} {\bibinfo  {journal} {Phys. Rev. A}\ }\textbf {\bibinfo {volume} {56}},\ \bibinfo {pages} {1163} (\bibinfo {year} {1997})}\BibitemShut {NoStop}%
\bibitem [{\citenamefont {Scarani}\ and\ \citenamefont {Gisin}(2001)}]{scarani}%
  \BibitemOpen
  \bibfield  {author} {\bibinfo {author} {\bibfnamefont {V.}~\bibnamefont {Scarani}}\ and\ \bibinfo {author} {\bibfnamefont {N.}~\bibnamefont {Gisin}},\ }\href {\doibase 10.1103/PhysRevA.65.012311} {\bibfield  {journal} {\bibinfo  {journal} {Phys. Rev. A}\ }\textbf {\bibinfo {volume} {65}},\ \bibinfo {pages} {012311} (\bibinfo {year} {2001})}\BibitemShut {NoStop}%
\bibitem [{\citenamefont {Chiribella}\ \emph {et~al.}(2013)\citenamefont {Chiribella}, \citenamefont {D'Ariano},\ and\ \citenamefont {Perinotti}}]{chiribella2013pra}%
  \BibitemOpen
  \bibfield  {author} {\bibinfo {author} {\bibfnamefont {G.}~\bibnamefont {Chiribella}}, \bibinfo {author} {\bibfnamefont {G.~M.}\ \bibnamefont {D'Ariano}}, \ and\ \bibinfo {author} {\bibfnamefont {P.}~\bibnamefont {Perinotti}},\ }\href@noop {} {\bibfield  {journal} {\bibinfo  {journal} {Physical Review A}\ }\textbf {\bibinfo {volume} {88}},\ \bibinfo {pages} {022318} (\bibinfo {year} {2013})}\BibitemShut {NoStop}%
\bibitem [{\citenamefont {Oreshkov}\ \emph {et~al.}(2012)\citenamefont {Oreshkov}, \citenamefont {Costa},\ and\ \citenamefont {Brukner}}]{brukner12}%
  \BibitemOpen
  \bibfield  {author} {\bibinfo {author} {\bibfnamefont {O.}~\bibnamefont {Oreshkov}}, \bibinfo {author} {\bibfnamefont {F.}~\bibnamefont {Costa}}, \ and\ \bibinfo {author} {\bibfnamefont {{\v{C}}.}~\bibnamefont {Brukner}},\ }\href {\doibase 10.1038/ncomms2076} {\bibfield  {journal} {\bibinfo  {journal} {Nature Communications}\ }\textbf {\bibinfo {volume} {3}},\ \bibinfo {pages} {1092} (\bibinfo {year} {2012})}\BibitemShut {NoStop}%
\bibitem [{\citenamefont {Chiribella}\ \emph {et~al.}(2009)\citenamefont {Chiribella}, \citenamefont {D'Ariano},\ and\ \citenamefont {Perinotti}}]{chiribella2009theoretical}%
  \BibitemOpen
  \bibfield  {author} {\bibinfo {author} {\bibfnamefont {G.}~\bibnamefont {Chiribella}}, \bibinfo {author} {\bibfnamefont {G.~M.}\ \bibnamefont {D'Ariano}}, \ and\ \bibinfo {author} {\bibfnamefont {P.}~\bibnamefont {Perinotti}},\ }\href@noop {} {\bibfield  {journal} {\bibinfo  {journal} {Physical Review A}\ }\textbf {\bibinfo {volume} {80}},\ \bibinfo {pages} {022339} (\bibinfo {year} {2009})}\BibitemShut {NoStop}%
\bibitem [{\citenamefont {D.et.al}(2018)}]{ebler2018enhanced}%
  \BibitemOpen
  \bibfield  {author} {\bibinfo {author} {\bibfnamefont {E.}~\bibnamefont {D.et.al}},\ }\href@noop {} {\bibfield  {journal} {\bibinfo  {journal} {Physical Review Letters}\ }\textbf {\bibinfo {volume} {120}},\ \bibinfo {pages} {120502} (\bibinfo {year} {2018})}\BibitemShut {NoStop}%
\bibitem [{\citenamefont {Zhao}\ \emph {et~al.}(2020)\citenamefont {Zhao}, \citenamefont {Ma}, \citenamefont {Ren}, \citenamefont {Feng},\ and\ \citenamefont {Zheng}}]{zhao2020quantum}%
  \BibitemOpen
  \bibfield  {author} {\bibinfo {author} {\bibfnamefont {X.}~\bibnamefont {Zhao}}, \bibinfo {author} {\bibfnamefont {X.}~\bibnamefont {Ma}}, \bibinfo {author} {\bibfnamefont {W.}~\bibnamefont {Ren}}, \bibinfo {author} {\bibfnamefont {Y.}~\bibnamefont {Feng}}, \ and\ \bibinfo {author} {\bibfnamefont {Y.}~\bibnamefont {Zheng}},\ }\href@noop {} {\bibfield  {journal} {\bibinfo  {journal} {Physical Review A}\ }\textbf {\bibinfo {volume} {102}},\ \bibinfo {pages} {040601} (\bibinfo {year} {2020})}\BibitemShut {NoStop}%
\bibitem [{\citenamefont {Mukhopadhyay}\ and\ \citenamefont {Pati}(2020)}]{akpati2020}%
  \BibitemOpen
  \bibfield  {author} {\bibinfo {author} {\bibfnamefont {C.}~\bibnamefont {Mukhopadhyay}}\ and\ \bibinfo {author} {\bibfnamefont {A.~K.}\ \bibnamefont {Pati}},\ }\href {\doibase 10.1088/2399-6528/abbd77} {\bibfield  {journal} {\bibinfo  {journal} {Journal of Physics Communications}\ }\textbf {\bibinfo {volume} {4}},\ \bibinfo {pages} {105003} (\bibinfo {year} {2020})}\BibitemShut {NoStop}%
\bibitem [{\citenamefont {Srivastava}\ \emph {et~al.}(2024)\citenamefont {Srivastava}, \citenamefont {Pati}, \citenamefont {Chakrabarty},\ and\ \citenamefont {Bhattacharya}}]{grover_switch}%
  \BibitemOpen
  \bibfield  {author} {\bibinfo {author} {\bibfnamefont {S.}~\bibnamefont {Srivastava}}, \bibinfo {author} {\bibfnamefont {A.~K.}\ \bibnamefont {Pati}}, \bibinfo {author} {\bibfnamefont {I.}~\bibnamefont {Chakrabarty}}, \ and\ \bibinfo {author} {\bibfnamefont {S.}~\bibnamefont {Bhattacharya}},\ }\href {http://iopscience.iop.org/article/10.1088/1751-8121/ad9efc} {\bibfield  {journal} {\bibinfo  {journal} {Journal of Physics A: Mathematical and Theoretical}\ } (\bibinfo {year} {2024})}\BibitemShut {NoStop}%
\bibitem [{\citenamefont {Rubino}\ \emph {et~al.}(2017)\citenamefont {Rubino}, \citenamefont {Rozema}, \citenamefont {Feix}, \citenamefont {Araújo}, \citenamefont {Zeuner}, \citenamefont {Procopio},\ and\ \citenamefont {Brukner}}]{rubino2017exp}%
  \BibitemOpen
  \bibfield  {author} {\bibinfo {author} {\bibfnamefont {G.}~\bibnamefont {Rubino}}, \bibinfo {author} {\bibfnamefont {L.~A.}\ \bibnamefont {Rozema}}, \bibinfo {author} {\bibfnamefont {A.}~\bibnamefont {Feix}}, \bibinfo {author} {\bibfnamefont {M.}~\bibnamefont {Araújo}}, \bibinfo {author} {\bibfnamefont {J.~M.}\ \bibnamefont {Zeuner}}, \bibinfo {author} {\bibfnamefont {L.~M.}\ \bibnamefont {Procopio}}, \ and\ \bibinfo {author} {\bibfnamefont {{\v{C}}.}~\bibnamefont {Brukner}},\ }\href@noop {} {\bibfield  {journal} {\bibinfo  {journal} {Science Advances}\ }\textbf {\bibinfo {volume} {3}},\ \bibinfo {pages} {e1602589} (\bibinfo {year} {2017})}\BibitemShut {NoStop}%
\bibitem [{\citenamefont {\emph{et al.}}(2024)}]{caonatureexp}%
  \BibitemOpen
  \bibfield  {author} {\bibinfo {author} {\bibfnamefont {J.~B.}\ \bibnamefont {\emph{et al.}}},\ }\href@noop {} {\bibfield  {journal} {\bibinfo  {journal} {Nat Rev Phys}\ }\textbf {\bibinfo {volume} {6}},\ \bibinfo {pages} {483} (\bibinfo {year} {2024})}\BibitemShut {NoStop}%
\bibitem [{\citenamefont {Str\"omberg}\ \emph {et~al.}(2023)\citenamefont {Str\"omberg}, \citenamefont {Schiansky}, \citenamefont {Peterson}, \citenamefont {Quintino},\ and\ \citenamefont {Walther}}]{PhysRevLett.131.060803}%
  \BibitemOpen
  \bibfield  {author} {\bibinfo {author} {\bibfnamefont {T.}~\bibnamefont {Str\"omberg}}, \bibinfo {author} {\bibfnamefont {P.}~\bibnamefont {Schiansky}}, \bibinfo {author} {\bibfnamefont {R.~W.}\ \bibnamefont {Peterson}}, \bibinfo {author} {\bibfnamefont {M.~T.}\ \bibnamefont {Quintino}}, \ and\ \bibinfo {author} {\bibfnamefont {P.}~\bibnamefont {Walther}},\ }\href {\doibase 10.1103/PhysRevLett.131.060803} {\bibfield  {journal} {\bibinfo  {journal} {Phys. Rev. Lett.}\ }\textbf {\bibinfo {volume} {131}},\ \bibinfo {pages} {060803} (\bibinfo {year} {2023})}\BibitemShut {NoStop}%
\bibitem [{\citenamefont {Yanamandra}\ \emph {et~al.}(2023)\citenamefont {Yanamandra}, \citenamefont {Srinidhi}, \citenamefont {Bhattacharya}, \citenamefont {Chakrabarty},\ and\ \citenamefont {Goswami}}]{indranilda}%
  \BibitemOpen
  \bibfield  {author} {\bibinfo {author} {\bibfnamefont {S.}~\bibnamefont {Yanamandra}}, \bibinfo {author} {\bibfnamefont {P.~V.}\ \bibnamefont {Srinidhi}}, \bibinfo {author} {\bibfnamefont {S.}~\bibnamefont {Bhattacharya}}, \bibinfo {author} {\bibfnamefont {I.}~\bibnamefont {Chakrabarty}}, \ and\ \bibinfo {author} {\bibfnamefont {S.}~\bibnamefont {Goswami}},\ }\href@noop {} {\bibfield  {journal} {\bibinfo  {journal} {arXiv:2310.04819}\ } (\bibinfo {year} {2023})}\BibitemShut {NoStop}%
\bibitem [{\citenamefont {Fuchs}(1996)}]{ref13}%
  \BibitemOpen
  \bibfield  {author} {\bibinfo {author} {\bibfnamefont {C.~A.}\ \bibnamefont {Fuchs}},\ }\href {https://arxiv.org/abs/quant-ph/9611010} {\enquote {\bibinfo {title} {Information gain vs. state disturbance in quantum theory},}\ } (\bibinfo {year} {1996}),\ \bibinfo {note} {preprint, quant-ph/9611010}\BibitemShut {NoStop}%
\bibitem [{\citenamefont {Nielsen}\ and\ \citenamefont {Chuang}(2000)}]{nielsen}%
  \BibitemOpen
  \bibfield  {author} {\bibinfo {author} {\bibfnamefont {M.~A.}\ \bibnamefont {Nielsen}}\ and\ \bibinfo {author} {\bibfnamefont {I.~L.}\ \bibnamefont {Chuang}},\ }\href@noop {} {\emph {\bibinfo {title} {Quantum Computation and Quantum Information}}},\ \bibinfo {edition} {10th}\ ed.\ (\bibinfo  {publisher} {Cambridge University Press},\ \bibinfo {address} {Cambridge},\ \bibinfo {year} {2000})\BibitemShut {NoStop}%
\bibitem [{\citenamefont {Bennett}\ \emph {et~al.}(1992)\citenamefont {Bennett}, \citenamefont {Bessette}, \citenamefont {Brassard}, \citenamefont {Salvail},\ and\ \citenamefont {Smolin}}]{Bennett1992}%
  \BibitemOpen
  \bibfield  {author} {\bibinfo {author} {\bibfnamefont {C.~H.}\ \bibnamefont {Bennett}}, \bibinfo {author} {\bibfnamefont {F.}~\bibnamefont {Bessette}}, \bibinfo {author} {\bibfnamefont {G.}~\bibnamefont {Brassard}}, \bibinfo {author} {\bibfnamefont {L.}~\bibnamefont {Salvail}}, \ and\ \bibinfo {author} {\bibfnamefont {J.}~\bibnamefont {Smolin}},\ }\href {\doibase 10.1007/BF00191318} {\bibfield  {journal} {\bibinfo  {journal} {Journal of Cryptology}\ }\textbf {\bibinfo {volume} {5}},\ \bibinfo {pages} {3} (\bibinfo {year} {1992})}\BibitemShut {NoStop}%
\bibitem [{\citenamefont {Ekert}\ \emph {et~al.}(1994)\citenamefont {Ekert}, \citenamefont {Huttner}, \citenamefont {Palma},\ and\ \citenamefont {Peres}}]{Privecy_amp}%
  \BibitemOpen
  \bibfield  {author} {\bibinfo {author} {\bibfnamefont {A.~K.}\ \bibnamefont {Ekert}}, \bibinfo {author} {\bibfnamefont {B.}~\bibnamefont {Huttner}}, \bibinfo {author} {\bibfnamefont {G.~M.}\ \bibnamefont {Palma}}, \ and\ \bibinfo {author} {\bibfnamefont {A.}~\bibnamefont {Peres}},\ }\href {\doibase 10.1103/PhysRevA.50.1047} {\bibfield  {journal} {\bibinfo  {journal} {Phys. Rev. A}\ }\textbf {\bibinfo {volume} {50}},\ \bibinfo {pages} {1047} (\bibinfo {year} {1994})}\BibitemShut {NoStop}%
\bibitem [{\citenamefont {Horodecki}\ \emph {et~al.}(1995)\citenamefont {Horodecki}, \citenamefont {Horodecki},\ and\ \citenamefont {Horodecki}}]{horodecki1995}%
  \BibitemOpen
  \bibfield  {author} {\bibinfo {author} {\bibfnamefont {R.}~\bibnamefont {Horodecki}}, \bibinfo {author} {\bibfnamefont {I.}~\bibnamefont {Horodecki}}, \ and\ \bibinfo {author} {\bibfnamefont {M.}~\bibnamefont {Horodecki}},\ }\href@noop {} {\bibfield  {journal} {\bibinfo  {journal} {Physics Letters A}\ }\textbf {\bibinfo {volume} {200}},\ \bibinfo {pages} {340} (\bibinfo {year} {1995})}\BibitemShut {NoStop}%
\bibitem [{\citenamefont {Deutsch}\ \emph {et~al.}(1996)\citenamefont {Deutsch}, \citenamefont {Ekert}, \citenamefont {Jozsa}, \citenamefont {Machiavello}, \citenamefont {Popescu},\ and\ \citenamefont {Sanpera}}]{qpa_des}%
  \BibitemOpen
  \bibfield  {author} {\bibinfo {author} {\bibfnamefont {D.}~\bibnamefont {Deutsch}}, \bibinfo {author} {\bibfnamefont {A.}~\bibnamefont {Ekert}}, \bibinfo {author} {\bibfnamefont {R.}~\bibnamefont {Jozsa}}, \bibinfo {author} {\bibfnamefont {C.}~\bibnamefont {Machiavello}}, \bibinfo {author} {\bibfnamefont {S.}~\bibnamefont {Popescu}}, \ and\ \bibinfo {author} {\bibfnamefont {A.}~\bibnamefont {Sanpera}},\ }\href {\doibase 10.1103/PhysRevLett.77.2818} {\bibfield  {journal} {\bibinfo  {journal} {Physical Review Letters}\ }\textbf {\bibinfo {volume} {77}},\ \bibinfo {pages} {2818} (\bibinfo {year} {1996})}\BibitemShut {NoStop}%
\bibitem [{\citenamefont {Hardy}(2007)}]{hardy2007towards}%
  \BibitemOpen
  \bibfield  {author} {\bibinfo {author} {\bibfnamefont {L.}~\bibnamefont {Hardy}},\ }\href@noop {} {\bibfield  {journal} {\bibinfo  {journal} {Journal of Physics A: Mathematical and Theoretical}\ }\textbf {\bibinfo {volume} {40}},\ \bibinfo {pages} {3081} (\bibinfo {year} {2007})}\BibitemShut {NoStop}%
\bibitem [{\citenamefont {et.al}(2018)}]{goswami2018indefinite}%
  \BibitemOpen
  \bibfield  {author} {\bibinfo {author} {\bibfnamefont {G.~K.}\ \bibnamefont {et.al}},\ }\href {\doibase 10.1103/PhysRevLett.121.090503} {\bibfield  {journal} {\bibinfo  {journal} {Phys. Rev. Lett.}\ }\textbf {\bibinfo {volume} {121}},\ \bibinfo {pages} {090503} (\bibinfo {year} {2018})}\BibitemShut {NoStop}%
\bibitem [{\citenamefont {Araújo}\ \emph {et~al.}(2014)\citenamefont {Araújo}, \citenamefont {Feix}, \citenamefont {Costa},\ and\ \citenamefont {Brukner}}]{araujo2014computational}%
  \BibitemOpen
  \bibfield  {author} {\bibinfo {author} {\bibfnamefont {M.}~\bibnamefont {Araújo}}, \bibinfo {author} {\bibfnamefont {A.}~\bibnamefont {Feix}}, \bibinfo {author} {\bibfnamefont {F.}~\bibnamefont {Costa}}, \ and\ \bibinfo {author} {\bibfnamefont {{\v{C}}.}~\bibnamefont {Brukner}},\ }\href@noop {} {\bibfield  {journal} {\bibinfo  {journal} {Physical Review Letters}\ }\textbf {\bibinfo {volume} {113}},\ \bibinfo {pages} {250402} (\bibinfo {year} {2014})}\BibitemShut {NoStop}%
\end{thebibliography}%

\end{document}